\documentclass{cimento}

\usepackage{graphicx}
\usepackage{amssymb}

\title{Searching for New Physics with Flavor Violating Observables}
\author{Wolfgang Altmannshofer\from{ins:fermilab}}
\instlist{\inst{ins:fermilab} Fermi National Accelerator Laboratory, P.O.~Box 500,
  Batavia, IL 60510, USA}

\PACSes{
\PACSit{12.60.-i}{Models beyond the standard model}
\PACSit{13.20.He}{Leptonic, semileptonic, and radiative decays of bottom mesons}
\PACSit{13.25.Ft}{Hadronic decays of charmed mesons}}

\begin{document}

\maketitle

\begin{abstract}
In this talk, I review the status and prospects of several low energy flavor observables that are highly sensitive to New Physics effects.
In particular I discuss the implications for possible New Physics in $b \to s$ transitions coming from the recent experimental results on the $B_s$ mixing phase, the branching ratio of the rare decay $B_s \to \mu^+ \mu^-$, and angular observables in the $B \to K^* \mu^+\mu^-$ decay.
Also the recent evidence for direct CP violation in singly Cabibbo suppressed charm decays and its interpretation in the context of New Physics models is briefly discussed.
\end{abstract}

\section{Introduction}

In the Standard Model (SM), flavor changing neutral current (FCNC) processes are absent at the tree level. They only appear at the loop level and are further strongly suppressed by small CKM mixing angles. Consequently, FCNCs are highly sensitive probes of any new physics (NP) that is not flavor blind. 
Remarkably, up to now all experimental results on flavor observables are consistent with SM expectations and lead to strong indirect constraints on NP models even at energy scales far beyond the direct reach of colliders. 
Indeed if new degrees of freedom exist that couple to quarks at tree level with generic flavor and CP violating  interactions, flavor constraints, in particular constraints from neutral Kaon mixing, require the corresponding NP scale to be above $\Lambda \gtrsim 10^4$~TeV.
If NP does exist at the TeV scale, as naturalness arguments indicate, then current flavor constraints already imply that its flavor structure has to be highly non-generic.

In this talk I describe the impact that the recent experimental progress on $B$ and charm physics observables has on models of NP.
After reviewing the status of the tension between $B \to \tau \nu$ and $\sin2\beta$ in sec.~\ref{sec:btaunu}, I discuss in sec.~\ref{sec:Bsmix} the current constraints on CP violation in $B_s$ mixing. In sec.~\ref{sec:Bsmm} and sec.~\ref{sec:BKstar} the implications of the recent experimental results on $B_s \to \mu^+\mu^-$ and $B \to K^* \mu^+ \mu^-$ for NP models are analyzed. Finally, in sec.~\ref{sec:ACP} I discuss the recent evidence for direct CP violation in singly Cabibbo suppressed charm decays and its interpretation in the context of NP models.

\section{\boldmath $B \to \tau \nu$ and $\sin2\beta$} \label{sec:btaunu}

Combining BaBar and Belle results of the $B \to \tau \nu$ decay leads to the following average for its branching ratio~\cite{Asner:2010qj}
\begin{equation}
 {\mathcal B}(B \to \tau \nu )_{\rm exp} = (1.64 \pm 0.34) 10^{-4} ~.
\end{equation}
In the SM, the charged current $B \to \tau \nu$ decay proceeds through tree level exchange of a $W$ boson and is helicity suppressed by the $\tau$ mass. The SM prediction of the branching ratio depends sensitively on the value of the CKM element $|V_{ub}|$. Using an average, obtained from direct measurements in inclusive and exclusive semi-leptonic B decays, $|V_{ub}| = (3.89\pm0.44) 10^{-3}$~\cite{Nakamura:2010zzi} as well as the precise lattice determination of the B meson decay constant $f_B = (190 \pm 4)$~MeV~\cite{Davies:2012qf}, one obtains
\begin{equation}
 {\mathcal B}(B \to \tau \nu )_{\rm SM} = (0.97 \pm 0.22) 10^{-4} ~.
\end{equation}
Within $2\sigma$, this is compatible with the experimental value.

If one instead uses an indirect determination of $|V_{ub}|$ from the measurements of $\sin2\beta$ as well as $\Delta M_d/\Delta M_s$ one finds
\begin{equation}
 {\mathcal B}(B \to \tau \nu )_{\rm SM} = (0.75 \pm 0.10) 10^{-4} ~,
\end{equation}
which is more than a factor of two and almost $3\sigma$ below the experimental value. Interpreted as a hint for NP, this tension between ${\mathcal B}(B \to \tau \nu)$ and $\sin2\beta$ can be addressed either by a sizable negative NP phase in $B_d$ mixing or by O(1) NP effects in $B \to \tau \nu$. A well known example is tree level charged Higgs exchange that can lead to large modifications of $B \to \tau \nu$. While in two Higgs doublet models of type II, the charged Higgs contribution necessarily interferes destructively with the SM, the more general framework of two Higgs doublet models with minimal flavor violation (MFV) allows also for constructive interference and an explanation of the tension~\cite{Blankenburg:2011ca}.

\section{\boldmath CP Violation in $B_s$ Mixing} \label{sec:Bsmix}

CP violation in $B_s$ mixing is strongly suppressed in the SM and therefore an excellent probe of NP. Assuming NP only in the dispersive part of the $B_s$ mixing amplitude $M_{12}$, a possible NP phase $\phi_s^{\rm NP}$ enters in a correlated way the semi-leptonic asymmetry in the decay of B mesons to ``wrong sign'' leptons
\begin{equation}
a_{\rm SL}^s = \frac{\Gamma(\bar B_s \to X \ell^+ \nu) - \Gamma(B_s \to X \ell^- \nu)}{\Gamma(\bar B_s \to X \ell^+ \nu) + \Gamma(B_s \to X \ell^- \nu)} = \left| \frac{\Gamma_{12}^s}{M_{12}^s}\right| \sin (\phi_s^{\rm SM} + \phi_s^{\rm NP}) 
\end{equation}
and the time dependent CP asymmetries in decays to CP eigenstates $f$
\begin{equation}
S_f \sin(\Delta M_s t) = \frac{\Gamma(\bar B_s(t) \to f) - \Gamma(B_s(t) \to f)}{\Gamma(\bar B_s(t) \to f) + \Gamma(B_s(t) \to f)}~, ~~ S_f = \sin(2|\beta_s| - \phi_s^{\rm NP}) ~.
\end{equation}
The small SM phases are $\phi_s^{\rm SM} \simeq 0.2^\circ$ and $\beta_s \simeq 1^\circ$.
A related observable is the like-sign dimuon charge asymmetry $A_{\rm SL}^b$ at D0. Assuming that it is caused by CP violation in B mixing, $A_{\rm SL}^b$ is a combination of the semileptonic asymmetries in the $B_s$ and $B_d$ system~\cite{Abazov:2011yk}
\begin{equation}
A_{\rm SL}^b \simeq 0.41 a_{\rm SL}^s + 0.59 a_{\rm SL}^d ~.
\end{equation}
The large value of $A_{\rm SL}^b = (-0.787\pm0.172\pm0.093)\%$ measured by D0~\cite{Abazov:2011yk} is almost $4\sigma$ above the tiny SM prediction.

\begin{figure}[t]
\centering
\includegraphics[width=0.4\textwidth]{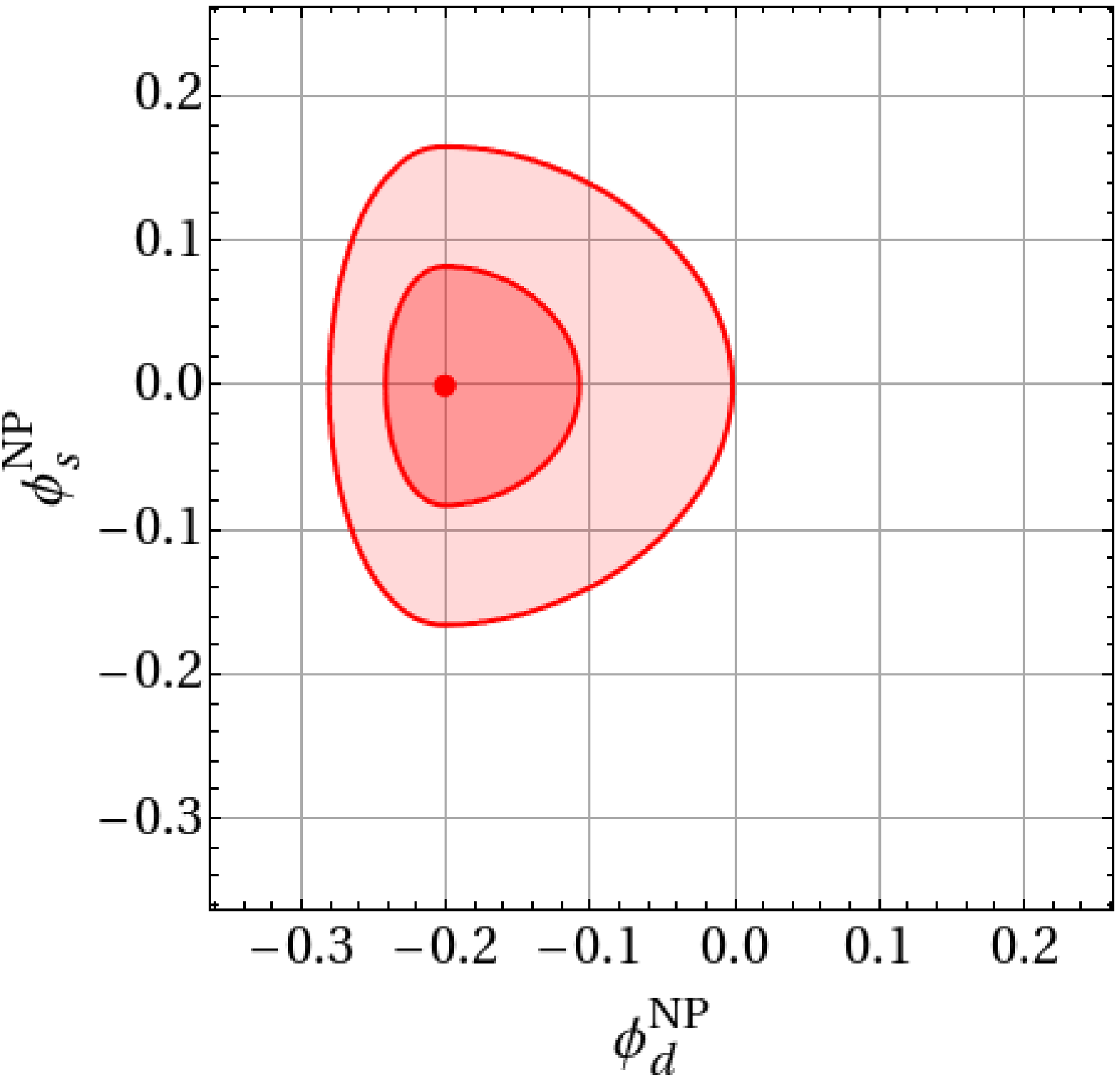} ~~~~~~~
\includegraphics[width=0.4\textwidth]{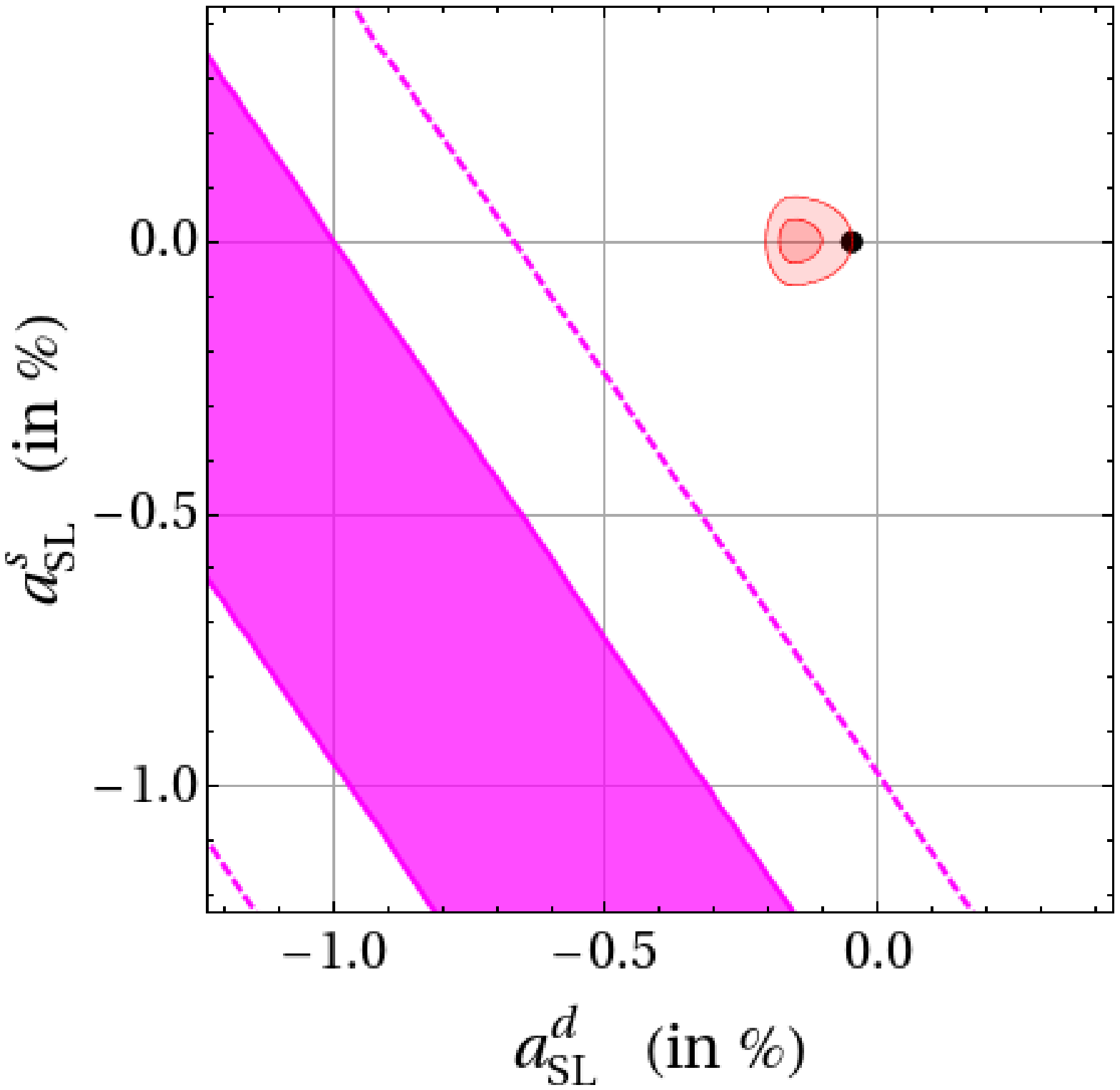}
\caption{
Left: Constraints on NP phases in $B_d$ and $B_s$ mixing from measurements of time-dependent CP asymmetries in $B_d \to \psi K_s$, $B_s \to \psi\phi$ and $B_s \to \psi \pi\pi$. Right: Corresponding region in the $a_{\rm SL}^d$ - $a_{\rm SL}^s$ plane. The black point shows the SM prediction, the diagonal band shows the measurement of $A_{\rm SL}^b$ from D0. (Update from~\cite{Altmannshofer:2011iv}.)
}
\label{fig:Bs_mixing}
\end{figure}

On the other hand, the latest experimental results on time dependent CP asymmetries, in particular the recent results from LHCb~\cite{LHCb-CONF-2012-002}, are all compatible with SM expectations. The left plot of Fig.~\ref{fig:Bs_mixing} shows the allowed ranges for NP phases in B meson mixing taking into account the most recent data on the time dependent CP asymmetry in $B_d \to \psi K_s$ from the B factories, the time dependent CP asymmetry in $B_s \to \psi\phi$ from CDF and D0 as well as the time dependent CP asymmetries in $B_s \to \psi\phi$ and $B_s \to \psi \pi\pi$ from LHCb. While the $B_s$ mixing phase is perfectly consistent with the SM expectation, the tensions in the Unitarity Triangle slightly prefer a small negative NP phase in $B_d$ mixing, $\phi_d^{\rm NP} = -0.2 \pm 0.1$. 

The right plot of Fig.~\ref{fig:Bs_mixing} shows the possible values for the semileptonic asymmetries $a_{\rm SL}^d$ and $a_{\rm SL}^s$, given the constraints on the NP phases. It is evident that the large like-sign dimuon charge asymmetry observed by D0 cannot be explained by NP in the dispersive part of the B mixing amplitude alone~\cite{Bobeth:2011st,Altmannshofer:2011iv,Lenz:2012az}.

The small preference for a negative NP phase in $B_d$ mixing on the other hand can be addressed for example in two Higgs doublet models with MFV through tree level exchange of flavor changing neutral Higgs bosons. If the quartic couplings in the Higgs potential are MSSM-like, these models predict a much larger NP phase in $B_s$ mixing than in $B_d$ mixing~\cite{Buras:2010mh}, which is clearly excluded by the current data. With more general Higgs potentials however, one generically finds $\phi_d^{\rm NP} \simeq \phi_s^{\rm NP}$~\cite{Buras:2010zm} and a small $\phi_d^{\rm NP}$ can still be accommodated. In the context of supersymmetric models with MFV, this is possible if the Higgs sector is extended by physics beyond the MSSM~\cite{Altmannshofer:2011rm,Altmannshofer:2011iv}.
These models will be challenged soon with improved measurements of the time-dependent CP asymmetries in $B_s \to \psi\phi$ and $B_s \to \psi \pi\pi$ by LHCb.

\section{\boldmath $B_s \to \mu^+ \mu^-$ and $B_d \to \mu^+ \mu^-$} \label{sec:Bsmm}

In the SM, the rare leptonic decays $B_s \to \mu^+ \mu^-$ and $B_d \to \mu^+ \mu^-$ are strongly helicity suppressed by the muon mass and their branching ratios are tiny -- at the level of $10^{-9}$ and $10^{-10}$, respectively.
Due to the remarkably precise determinations of the B meson decay constants on the lattice~\cite{Davies:2012qf}, the SM predictions for both decays have reached theory uncertainties of less than $10\%$. Taking into account the correction to $\mathcal{B}(B_s \to \mu^+ \mu^-)$ coming from the large width difference in the $B_s$ meson system recently pointed out in~\cite{deBruyn:2012wj,deBruyn:2012wk}, one gets
\begin{equation} \label{eq:Bqmm_SM}
 {\mathcal B}(B_s \to \mu^+ \mu^-)_{\rm SM} = (3.32 \pm 0.17)10^{-9} ~,~ {\mathcal B}(B_d \to \mu^+ \mu^-)_{\rm SM} = (1.0 \pm 0.1)10^{-10}~.
\end{equation}
In extensions of the SM where NP contributions to these decays arise from scalar operators, e.g. through neutral Higgs exchange in the MSSM with large $\tan\beta$, the helicity suppression is lifted and order of magnitude enhancements of the branching ratios are possible. On the other hand, if the NP induces operators with a helicity structure analogous to the SM, e.g. through $Z$ or $Z^\prime$ exchange, it has been shown that constraints from the semileptonic decays $B \to X_s \ell^+\ell^-$ and $B \to K^* \mu^+ \mu^-$ only allow an enhancement of ${\mathcal B}(B_s \to \mu^+ \mu^-)$ up to $5.6 \times 10^{-9}$~\cite{Altmannshofer:2011gn}.

On the experimental side, D0~\cite{Abazov:2010fs}, Atlas~\cite{Aad:2012pn}, CMS~\cite{Chatrchyan:2012rg} and LHCb~\cite{Aaij:2012ac} give upper bounds on the branching ratio. CDF reported an excess in $B_s \to \mu^+ \mu^-$ candidates leading to a two sided limit on $\mathcal{B}(B_s \to \mu^+\mu^-)$ at 95\% C.L.~\cite{Aaltonen:2011fi} that, at the 2$\sigma$ level, is consistent with the upper bounds. For $B_d \to \mu^+ \mu^-$, upper bounds are reported by CDF~\cite{Aaltonen:2011fi}, CMS~\cite{Chatrchyan:2012rg} and LHCb~\cite{Aaij:2012ac}. Performing a naive combination of the available results one finds
\begin{equation} \label{eq:Bqmm_exp}
\mathcal{B}(B_s \to \mu^+ \mu^-)_{\rm exp} = (2.4 \pm 1.6) 10^{-9} ~,~ \mathcal{B}(B_d \to \mu^+ \mu^-)_{\rm exp} < 8.6 \times 10^{-10} ~.
\end{equation}
While in $B_d \to \mu^+ \mu^-$ there is still room for almost an order of magnitude enhancement, the recent results on $B_s \to \mu^+ \mu^-$ only allow for moderate deviations from the SM prediction. On the one hand this leads to very strong constraints on models with scalar operators, on the other hand, models without scalar operators are starting to be probed by $B_s \to \mu^+ \mu^-$ only now.

An important observable in the future will be the ratio of the $B_s$ and $B_d$ branching ratios. In models with minimal flavor violation there exists a strong correlation
\begin{equation}
\frac{{\mathcal{B}}(B_s \to \mu^+ \mu^-)}{{\mathcal B}(B_d \to \mu^+ \mu^-)} \simeq \frac{f_{B_s}^2}{f_{B_d}^2} \frac{\tau_{B_s}}{\tau_{B_d}} \frac{|V_{ts}|^2}{|V_{td}|^2} \simeq 35 ~.
\end{equation}
Given the existing bounds on ${\mathcal B}(B_s \to \mu^+ \mu^-)$, an enhancement of ${\mathcal B}(B_d \to \mu^+ \mu^-)$ by more than a factor of 2 would not only be a clear indication of NP, but also of new sources of flavor violation beyond the CKM matrix.

\section{\boldmath Angular Observables in $B \to K^* \mu^+ \mu^-$} \label{sec:BKstar}

The semileptonic exclusive $B \to K^* (\to K^+ \pi^-) \mu^+ \mu^-$ decay and its conjugated mode $\bar B \to \bar K^* (\to K^- \pi^+) \mu^+ \mu^-$ are described by 4-fold differential decay distributions $d\Gamma$ and $d \bar\Gamma$ in terms of the dimuon invariant mass $q^2$, and three angles $\theta_{K^*}$, $\theta_\ell$ and $\phi$ (see {\it e.g.}~\cite{Altmannshofer:2008dz} for details), offering a multitude of observables that can be used to search for NP.

One dimensional angular distributions give access to the well known observables $F_L$, the $K^*$ longitudinal polarization fraction, and $A_{\rm FB}$, the forward-backward asymmetry.
Also the transversal asymmetry $S_3$ and the T-odd CP asymmetry $A_9$ can be obtained from a one dimensional angular analysis
\begin{eqnarray}
\frac{d(\Gamma + \bar \Gamma)}{dq^2 d\cos\theta_{K^*}} &\propto & 2 F_L \cos^2\theta_{K^*} + (1-F_L) \sin^2\theta_{K^*} ~, \\
\frac{d(\Gamma - \bar \Gamma)}{dq^2 d\cos\theta_\ell} &\propto & A_{\rm FB} \cos\theta_\ell +  \frac{3}{4} F_L \sin^2\theta_\ell + \frac{3}{8} (1-F_L) (1 + \cos^2\theta_\ell) ~, \\ 
\frac{d(\Gamma + \bar \Gamma)}{dq^2 d\phi} &\propto&  1 + S_3 \cos2\phi + A_9 \sin2\phi ~.
\end{eqnarray}
Additional observables, like the T-odd CP asymmetries $A_7$ and $A_8$ require two or three dimensional angular analyses~\cite{Bobeth:2008ij}.

Measurements of angular observables exist from BaBar~\cite{talk}, Belle~\cite{Wei:2009zv}, CDF~\cite{Aaltonen:2011ja} and LHCb~\cite{LHCb-CONF-2012-008} and all show reasonable agreement with the SM predictions so far.  
The various observables each depend in a characteristic way on the Wilson coefficients of the dimension six $\Delta F=1$ operators contributing to $B \to K^* \mu^+ \mu^-$.
Particularly interesting are the observables $S_3$ and $A_9$ as they are easily accessible, very small in the SM, and highly sensitive to CP conserving and CP violating right-handed currents that are predicted in various NP models. 

\begin{figure}[t]
\centering
\includegraphics[width=\textwidth]{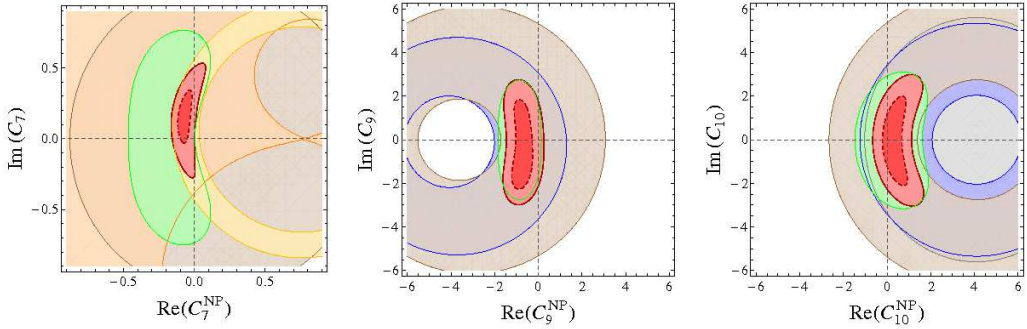}
\caption{
Allowed regions in the complex planes of the Wilson coefficients $C_{7,9,10}$ at 1 and 2$\sigma$ (red). Shown are also the individual 2$\sigma$ constraints from ${\mathcal B}(B_s \to \mu^+\mu^-)$ (gray), ${\mathcal B}(B \to X_s \ell^+\ell^-)$ (brown), ${\mathcal B}(B \to K \mu^+\mu^-)$ (blue), $B \to K^* \mu^+\mu^-$ (green), ${\mathcal B}(B \to X_s \gamma)$ (yellow) and $A_{\rm CP}(B \to X_s \gamma)$ (orange). (From~\cite{ASupdate}.)
}
\label{fig:WCconstraints}
\end{figure}

Analyses that put constraints on the Wilson coefficients and therefore determine in a model independent way the room left for NP in $B \to K^* \mu^+ \mu^-$ have been performed {\it e.g.} for the magnetic penguin operators~\cite{DescotesGenon:2011yn}, the SM operator basis~\cite{Bobeth:2011nj,Beaujean:2012uj} and for the very general case of the SM operators as well as their chirality flipped counterparts~\cite{Altmannshofer:2011gn,ASupdate}. In~\cite{ASupdate} we include all the available experimental results on the $B \to K^* \mu^+ \mu^-$ observables as well as all relevant observables in the $B \to K \mu^+\mu^-$, $B\to X_s \ell^+\ell^-$, $B \to X_s \gamma$, $B \to K^* \gamma$ and $B_s \to \mu^+\mu^-$ decays. Results for the Wilson coefficients $C_{7,9,10}$ are shown in Fig.~\ref{fig:WCconstraints}.
Analogous results for the right-handed coefficients $C^\prime_{7,9,10}$ can be found in~\cite{ASupdate}.

The strongest bounds currently come from branching ratios and CP averaged angular coefficients, and therefore the imaginary parts of $C_{7,9,10}$ are much less constrained than the real parts. Future improved measurements of CP asymmetries in $B \to K^* \mu^+ \mu^-$ that are directly sensitive to new CP violating phases will be essential to probe the imaginary parts of the Wilson coefficients.

\section{\boldmath Direct CP Violation in $D \to K^+ K^-$ and $D \to \pi^+ \pi^-$} \label{sec:ACP}

CP Violation in the charm sector is highly Cabibbo suppressed in the SM and therefore very sensitive to possible NP effects. Current bounds on CP violating parameters in $D^0 - \bar D^0$ mixing are at the level of $10\% - 20\%$~\cite{Asner:2010qj} and still far above the naive SM expectation of $O(|V_{ub}V_{cn}^*|/|V_{us}V_{cs}^*|) \simeq 10^{-3}$.
Naive SM estimates for direct CP violation in singly Cabibbo suppressed $D$ decays such as $D \to K^+ K^-$ and $D \to \pi^+ \pi^-$ are even smaller, $O(|V_{ub}V_{cb}^*|/|V_{us}V_{cs}^*| \alpha_s/\pi) \simeq 10^{-4}$.

Remarkably, the LHCb collaboration recently found first evidence for charm CP violation~\cite{Aaij:2011in}. The reported value for $\Delta A_{\rm CP}$, the difference in the time integrated CP asymmetries in $D \to K^+ K^-$ and $D \to \pi^+\pi^-$, is non-zero at $3.5\sigma$. This result has been confirmed by CDF~\cite{CDF:10784} and a combination that includes also previous results from BaBar and Belle leads to~\cite{Asner:2010qj}
\begin{equation}
 \Delta A_{\rm CP} = -(0.656 \pm 0.154)\% ~,
\end{equation}
which is approximately $4\sigma$ away from 0.
To a very good approximation, $\Delta A_{\rm CP}$ corresponds to the difference in the direct CP asymmetries in $D \to K^+ K^-$ and $D \to \pi^+ \pi^-$ and is therefore expected to be at least one order of magnitude smaller in the SM, unless the hadronic matrix elements entering the SM prediction are strongly enhanced with respect to naive estimates~\cite{Golden:1989qx}. While the required strong enhancement is not expected to be natural in the SM~\cite{Cheng:2012wr,Franco:2012ck}, it cannot be excluded presently~\cite{Brod:2011re,Bhattacharya:2012ah,Feldmann:2012js,Brod:2012ud}. Nonetheless, the interpretation of the experimental results as a signal of NP is interesting and motivated. Possible NP explanations typically will predict non-standard signals also in other low energy flavor observables or characteristic signatures at colliders that can be searched for and that can be used to test the NP hypothesis. 

Both model independent analyses~\cite{Isidori:2011qw}, and studies of concrete NP scenarios (see {\it e.g.}~\cite{Grossman:2006jg,Hochberg:2011ru,Giudice:2012qq,Altmannshofer:2012ur,Hiller:2012wf,KerenZur:2012fr}) show that NP explanations of large direct CP violation in singly Cabibbo suppressed $D^0$ decays are often highly constrained by other flavor observables, in particular $D^0 - \bar D^0$ mixing and $\epsilon^\prime / \epsilon$, {\it i.e.} direct CP violation in neutral Kaon decays.
Among the numerous NP possibilities, the one that generically avoids both these constraints are loop induced chromomagnetic dipole operators 
\begin{equation}
 O_8 = \frac{g_s}{16\pi^2} m_c ~\bar u (\sigma G) P_R c ~,~ O_8 = \frac{g_s}{16\pi^2} m_c ~\bar u (\sigma G) P_L c ~,
\end{equation}
that can lead to chirally enhanced contributions to the direct CP asymmetries. Studies of CP asymmetries in radiative $D \to V \gamma$ decays could help to probe the possible dipole origin of $\Delta A_{\rm CP}$~\cite{Isidori:2012yx}.

Among NP models that contribute significantly to $\Delta A_{\rm CP}$ through tree-level induced four fermion operators, only few are viable. Known examples include models where the NP contribution is mediated by scalars with masses at the electro-weak scale and flavor changing couplings~\cite{Hochberg:2011ru,Altmannshofer:2012ur}. If these models induce $\Delta I = 3/2$ operators, they can be tested using isospin sum rules for CP asymmetries~\cite{Grossman:2012eb}.
The necessarily light scalars can be searched for at the LHC.

Finally, depending on the exact flavor structure of the tree-level models, $\Delta A_{\rm CP}$ can originate from either $A_{\rm CP}^{K^+K^-}$ or $A_{\rm CP}^{\pi^+\pi^-}$ or from both. This is in contrast to the dipole operator explanation where one (naively) expects $A_{\rm CP}^{K^+K^-} \simeq - A_{\rm CP}^{\pi^+\pi^-}$. Correspondingly, the separate measurement of the direct CP asymmetries in $D \to K^+K^-$ and $D \to \pi^+\pi^-$ would add valuable information to pin down the possible origin of direct CP violation in charm decays.

\section{Summary}

In absence of any direct evidence for new physics at the LHC, flavor physics continues to stay at the forefront of the search for new phenomena at the TeV scale.
While the current experimental results on rare decays like $B_s \to \mu^+\mu^-$ and $B \to K^* \mu^+\mu^-$ are in reasonable agreement with SM predictions, they still allow for sizable new physics effects, that can be uncovered with improved precision in the near future.
Moreover, anomalies in the current flavor data, like the tension between $B \to \tau \nu$ and $\sin2\beta$, the large dimuon charge asymmetry or the recent evidence for direct CP violation in charm decays might already be the first indirect signs of physics beyond the Standard Model.

Improved experimental results on all these observables will be most exciting and certainly improve our understanding of possible new physics at the TeV scale and beyond.  

\acknowledgments

I would like to thank the organizers for the invitation to this wonderful conference.
Fermilab is operated by Fermi Research Alliance, LLC under Contract No. De-AC02-07CH11359 with the United States Department of Energy.


\end{document}